\documentclass[pra,showpacs,twocolumn,superscriptaddress]{revtex4-1}
\usepackage{mathrsfs}
\usepackage{amsfonts}
\usepackage{amsmath}
\usepackage{txfonts}
\usepackage{amssymb}
\usepackage{graphicx}
\usepackage{bm}
\usepackage{color}
\usepackage[normalem]{ulem}

\newcommand{\ket}[1]{|#1\rangle}
\newcommand{\bra}[1]{\langle #1|}

\begin{document}
\title{Dark path holonomic qudit computation}
\author{Tomas Andr\'e}
\affiliation{Department of Physics and Astronomy, Uppsala University,
Box 516, Se-751 20 Uppsala, Sweden}
\author{Erik Sj\"oqvist}
\email{erik.sjoqvist@physics.uu.se}
\affiliation{Department of Physics and Astronomy, Uppsala University,
Box 516, Se-751 20 Uppsala, Sweden}
\date{\today}
\begin{abstract}
Non-adiabatic holonomic quantum computation is a method used to implement 
high-speed quantum gates with non-Abelian geometric phases associated with paths 
in state space. Due to their noise tolerance,  these phases can be used to construct 
error resilient quantum gates. We extend the holonomic dark path qubit scheme in 
[M.-Z. Ai {\it et al.}, Fundam. Res. {\bf 2}, 661 (2022)] to qudits. Specifically, we 
demonstrate one- and two-qudit universality by using the dark path technique. Explicit 
qutrit ($d=3$) gates are demonstrated and the scaling of the number of loops with the 
dimension $d$ is addressed. This scaling is linear and we show how any diagonal 
qudit gate can be implemented efficiently in any dimension.
\end{abstract}
\maketitle
\date{\today}

\section{Introduction}
The most common form of quantum computation is the circuit model, which is analogous 
to the circuits used 
for classical computers. Gates are replaced by unitary transformations (quantum gates) and 
bits by qubits. To achieve the computational advantage it is important to construct robust, 
noise-resilient quantum gates. A candidate for this is holonomic quantum computation 
\cite{zanardi99,sjoqvist12}, which is based on non-Abelian (matrix-valued) geometric 
phases in adiabatic \cite{wilczek84} or non-adiabatic \cite{anandan88} evolution.  
Such holonomic gates are only dependent on the geometry of the system's state space 
and thus are resilient to local errors in the quantum evolution. Recent theoretical and 
experimental advances in holonomic quantum computation can be found in 
Refs.~\cite{zhang18,xu18a,liu19,li20,dong21a,chen21,dong21b,setiawan21,alves22} 
and \cite{xu18b,huang19,egger19,xu20,dong21c,xu21a,wu22,ai22}, respectively.

The idea that elements of computation should be limited to qubits is sort 
of an arbitrary choice that most likely rose out of convenience due to binary logic. So why 
binary logic? It is simply the easiest non-trivial example: in binary logic, things can be either 
$0$ or $1$, {\tt True} or {\tt False}, {\bf on} or {\bf off}, etc. Due to its simplicity, it is no wonder 
that this is how the first computer was designed. But are we limited to bits? As early as 
1840, a mechanical trinary (three-valued logic) calculation device was built by Fowler 
\cite{glusker05}, and in 1958 the first electronic trinary computer was developed by the Soviet 
Union \cite{bursentsov11}. Although the trinary computer had many advantages over the 
binary one, it never saw the same widespread success. However, there is nothing in theory that 
forbids a higher dimensional computational basis, even more so when it comes to quantum 
computers, where the implementation of the elements of computation already surpasses 
the simplicity of {\bf on} and {\bf off}. Indeed, qudits have been implemented experimentally 
and shown definite advantages \cite{lanyon09,baekkegaard19} as well as paving the way 
for achieving large-scale quantum computation \cite{chi22}. Thus, one may consider $d$ 
dimensional `qudits' as primitive units of quantum information, with promising results that 
show potential, some of them reviewed in Ref.~\cite{wang20}.

Here, we develop a qudit generalization of the idea of dark paths proposed in Ref.~\cite{ai22} 
for implementing non-adiabatic holonomic qubit gates. By this, we combine the advantages of 
improved robustness associated with the dark path approach with the enlarged encoding 
space and improved gate efficiency of higher-dimensional quantum information 
units \cite{wang20}. In the next two sections, we extend the dark path idea to the 
qutrit ($d=3$). We examine the robustness of the $d=3$ gates to systematic errors 
in the Rabi frequencies of the laser induced transitions. The case of general $d$ is 
examined in Secs.~\ref{sec:gen} and \ref{sec:2-qudit}. The paper ends with the conclusions. 

\section{Dark path setting}
\begin{figure}[h!]
\centering
\includegraphics[width=0.45\textwidth]{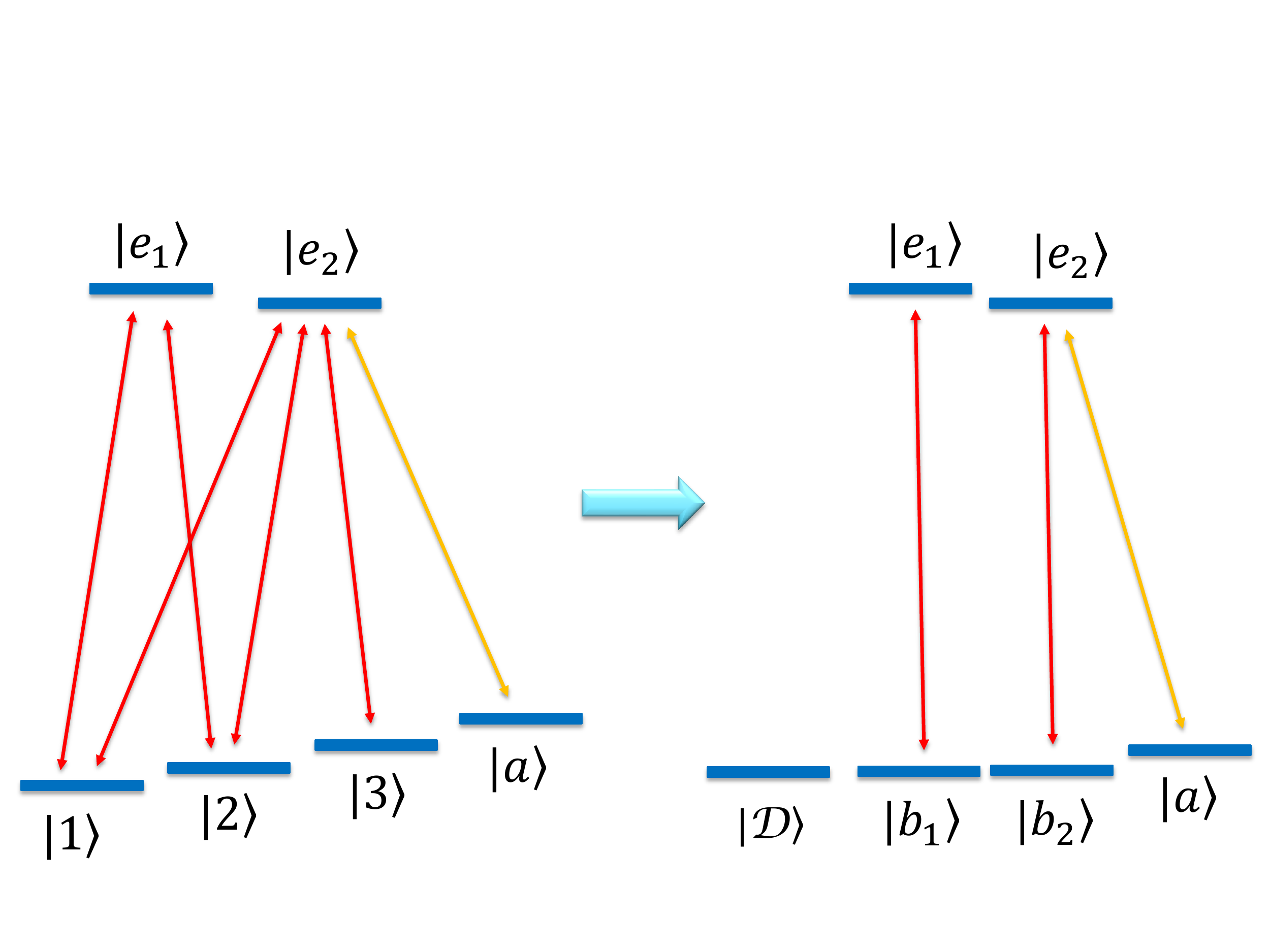}
\caption{Level setting for realizing a dark path holonomic qutrit gate (left panel). The 
specific coupling structure may be realized by inducing transitions between hyperfine levels 
in trapped ions. It allows for a Morris-Shore transformation \cite{morris83} 
applied to the qutrit levels $\ket{1},\ket{2},\ket{3}$, resulting in a single dark energy eigenstate 
$\ket{\mathcal{D}}$ and two bright states $\ket{b_1},\ket{b_2}$, while leaving the auxiliary state 
$\ket{a}$ untouched (right panel).}
\label{fig:setting1}
\end{figure}

The key point of the qutrit dark path setting is to look for a level structure with three 
ground states $\ket{k}$, $k=1,2,3$, encompassing a single fixed (time 
independent) dark eigenstate, as well as an extra fourth auxiliary lower level $\ket{a}$. 
As the number of dark eigenstates in the computational qutrit subspace ${\rm Span} 
\{ \ket{1},\ket{2},\ket{3} \}$ equals the difference between the number of excited states 
and the number of ground states \cite{shkolnikov20}, this amounts to coupling the 
ground state levels to two  excited states $\ket{e_1},\ket{e_2}$. The desired coupling 
structure is described by the Hamiltonian (see left panel of Fig.~\ref{fig:setting1})
\begin{eqnarray}
\label{eq:Ham}
H^{(3)} = \sum_{k = 1}^3 \sum_{l =1}^{2} \omega_{k,l}\ket{k}\bra{e_l}  + 
\frac{\Omega_{a}(t)}{2}\ket{a}\bra{e_2}  + \text{H.c.} 
\end{eqnarray}
with $\omega_{3,1}=0$. Laser-induced dipole transitions between hyperfine levels of 
trapped ions provide an ideal platform for realizing such coupling structures \cite{low20}. 
The specific Hamiltonian $H^{(3)}$ allows for a Morris-Shore transformation \cite{morris83} 
applied to the qutrit levels $\ket{1},\ket{2},\ket{3}$ 
only, yielding (see right panel of Fig.~\ref{fig:setting1})
\begin{eqnarray}
\label{eq:Ham_d}
H^{(3)} = \sum_{k = 1}^2 \frac{\Omega_k(t)}{2}e^{-i\phi_k}\ket{b_k}\bra{e_k}  + 
\frac{\Omega_a(t)}{2}\ket{a}\bra{e_2}  +\,\text{H.c.}
\end{eqnarray}
with $\Omega_k$ being real-valued time dependent Rabi frequencies and $\phi_k$
time independent phases, by employing the dark-bright basis 
\begin{eqnarray}
\label{eq:states}
\ket{\mathcal{D}} & = & \cos\theta\ket{1} + e^{i\chi}\sin\theta\cos\varphi\ket{2} + 
e^{i\xi}\sin\theta\sin\varphi\ket{3},
\nonumber \\
\ket{b_1} & = & \frac{1}{\sqrt{1-\sin^2\theta\sin^2\varphi}} 
\left(-e^{-i\chi}\sin\theta\cos\varphi\ket{1} + \cos\theta\ket{2} \right),
\nonumber \\
\ket{b_2} & = & \frac{1}{\sqrt{1-\sin^2\theta\sin^2\varphi}} 
\Big( \sin \theta \cos \theta \sin\varphi\ket{1} 
\nonumber \\ 
 & & + e^{i\chi} \sin^2 \theta \sin \varphi \cos \varphi \ket{2} + 
 e^{i\xi}(\sin^2\theta\sin^2\varphi - 1)\ket{3} \Big) , 
 \nonumber \\ 
\end{eqnarray}
where $\ket{\mathcal{D}}$ is a dark energy eigenstate satisfying $H^{(3)} \ket{\mathcal{D}} = 0$. 
Note that $\ket{\mathcal{D}},\ket{b_1},\ket{b_2}$ span the computational subspace, i.e., 
${\rm Span} \{ \ket{\mathcal{D}},\ket{b_1},\ket{b_2} \} = {\rm Span} \{ \ket{1},\ket{2},\ket{3} \}$. 
The original $\omega_{k,l}$ can be expressed in the parameters $\theta,\varphi,\chi,\xi$ by 
expanding the bright states on the right-hand side of Eq.~(\ref{eq:Ham_d}) in terms of the 
original qutrit levels $\ket{k}$ and by comparing with Eq.~(\ref{eq:Ham}).

A pair of dark path states $\ket{D_1(t)},\ket{D_2(t)}$ can now be defined. These states 
should satisfy two conditions: (i) they should be orthogonal to $\ket{\mathcal{D}}$, and 
(ii) their average energy $\bra{D_k(t)}H_d\ket{D_k(t)}$, $k=1,2,$ should vanish along their 
evolution paths. Explicitly, one may check that 
\begin{eqnarray}
\label{eq:dpaths}
\ket{D_1(t)} & = & \cos u(t) e^{-i\phi_1}\ket{b_1} + i\sin u(t) \ket{e_1},
\nonumber \\
\ket{D_2(t)} & = & \cos u(t)\cos v(t) e^{-i\phi_2}\ket{b_2} - i\sin u(t) \ket{e_2}
\nonumber \\ 
 & & - \cos u(t)\sin v(t) \ket{a} 
\end{eqnarray}
satisfy the dark path conditions. In fact, these states even satisfy the stronger condition  
\cite{sjoqvist12} $\bra{D_1 (t)}H_d\ket{D_2 (t)} = 0$, which opens up for holonomic 
gates provided the parameters $u(t),v(t)$ are chosen such that the qutrit subspace 
${\rm Span} \{ \ket{\mathcal{D}},  \ket{D_1 (t)},  \ket{D_2 (t)} \}$ evolves in a cyclic 
manner, i.e., that ${\rm Span} \{ \ket{\mathcal{D}},  \ket{D_1 (\tau)},  \ket{D_2 (\tau)} \} = 
{\rm Span} \{ \ket{1},  \ket{2},  \ket{3} \}$ for some run time $\tau$. This is achieved 
provided $u(\tau) = u(0) = v(\tau) = v(0) = 0$ so that each dark path starts in the 
respective bright state and travels along a curve and returns to the same bright state 
at $t=\tau$. 

\begin{figure}[h!]
\centering
\includegraphics[width=0.5\textwidth]{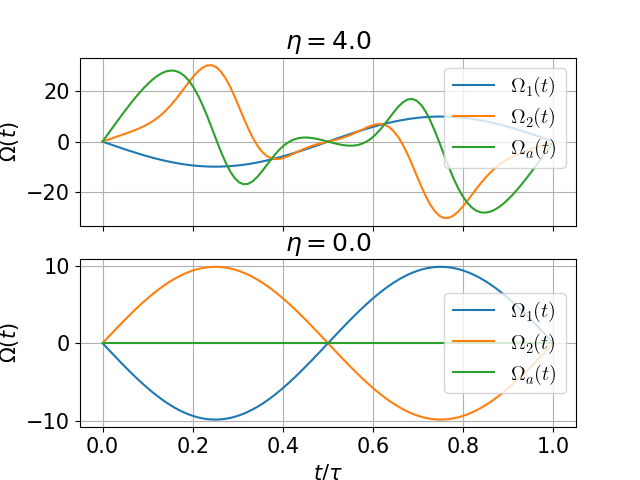}
\caption{Rabi frequencies in the qutrit scheme for zero and non-zero value of 
$\eta$. Note that only $\Omega_2(t)$ and $\Omega_a(t)$ change shape 
with $\eta$. This extends to the higher dimensional case in that only the pulse 
with the largest index and $\Omega_a(t)$ depend on $\eta$ for arbitrary $d$ [see 
Eq.~(\ref{eq:genRabi}) below].} 
\label{fig:rabi}
\end{figure}

One may now use the Schr{\"o}dinger equation to reverse engineer the time dependent 
parameters $\Omega_l(t)$. A calculation yields 
\begin{eqnarray}
\Omega_1(t) &=& -2\dot{u}(t),
 \nonumber \\ 
\Omega_2(t) &=& 2\left(\dot{v}(t)\cot u(t)\sin v(t) + \dot{u}(t)\cos v(t) \right),
\nonumber \\
\Omega_a(t) &=& 2\left(\dot{v}(t)\cot u(t)\cos v(t) - \dot{u}(t)\sin v(t) \right).
\end{eqnarray}
We follow Ref.~\cite{ai22} and choose 
\begin{eqnarray}
u(t) & = & \frac{\pi}{2}\sin^2\frac{\pi t}{\tau}, 
\nonumber \\ 
v(t) & = & \eta\left[1 - \cos u(t)\right] , 
\end{eqnarray}
which ensures cyclic evolution.
The parameter $\eta$ represents the coupling strength to the auxiliary state  
in the sense that $\ket{D_2 (t)}$ becomes independent of $\ket{a}$ when 
$\eta = 0$. The shape of the Rabi frequencies with this choice of $u$ and $v$ 
are displayed in Fig.~\ref{fig:rabi}. This completes the dark path construction in 
the qutrit case. 

\section{Holonomic dark path one-qutrit gates}
\subsection{Gate construction}
The multi-pulse single-loop setting \cite{herterich16} is used for dark path qutrit gates, 
thereby the loop is divided into two path segments by applying laser pulses. Explicitly, 
the pulses take the subspace ${\rm Span} \{ \ket{\mathcal{D}},\ket{b_1},\ket{b_2} \}$ into 
${\rm Span} \{ \ket{\mathcal{D}},\ket{e_1},\ket{e_2}\}$ and back, by applying phase shifts 
$\gamma_1,\gamma_2$ to the second segment relative to the first one. Note that 
$u\left(\frac{\tau}{2}\right) = \frac{\pi}{2}$, which implies that the duration is $\frac{\tau}{2}$ 
for both path segments. This results in the unitaries    
\begin{eqnarray}
U\left( \frac{\tau}{2},0 \right) & = & 
\ket{\mathcal{D}}\bra{\mathcal{D}} -i\big( \ket{e_1}\bra{b_1} +  \ket{b_1}\bra{e_1} \big) 
\nonumber \\ 
 & & - i\big( \ket{e_2}\bra{b_2} + \ket{b_2}\bra{e_2} \big) ,
\nonumber \\
U\left( \tau,\frac{\tau}{2} \right)  & = & \ket{\mathcal{D}}\bra{\mathcal{D}} +i e^{i\gamma_1} 
\big( \ket{b_1}\bra{e_1} +  \ket{e_1}\bra{b_1} \big) 
\nonumber \\ 
 & & + ie^{i\gamma_2}\big( \ket{b_2}\bra{e_2} +  \ket{e_2}\bra{b_2} \big) 
\end{eqnarray}
restricted to the part with non-trivial action on the computational subspace. 
By combining these unitaries, we obtain the one-qutrit gate 
\begin{eqnarray}
\label{eq:trit-gate-1-loop}
{\rm U}_3^{(1)}  & = & U\left( \tau,\frac{\tau}{2} \right) U\left( \frac{\tau}{2},0 \right) 
\nonumber \\
 & = & \ket{\mathcal{D}}\bra{\mathcal{D}} + 
e^{i\gamma_1} \ket{b_1}\bra{b_1} + e^{i\gamma_2}\ket{b_2}\bra{b_2}.
\end{eqnarray}

\begin{figure}[h!]
\centering
\includegraphics[width=0.45\textwidth]{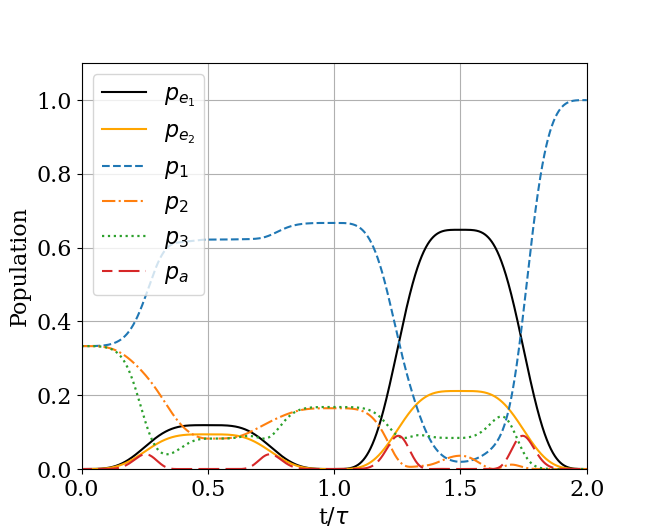}
\includegraphics[width=0.45\textwidth]{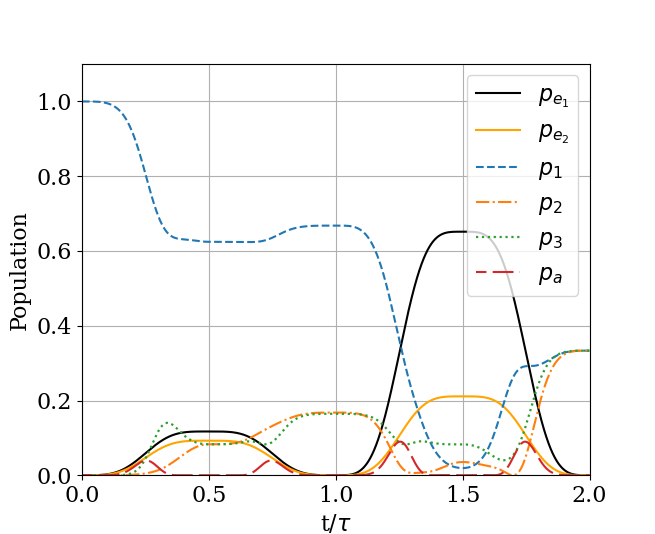}
\caption{The effect of the H$_3$-gate on the initial states $\frac{1}{\sqrt{3}} \big( \ket{1} + 
\ket{2} + \ket{3} \big)$ (upper panel) and $\ket{1}$ (lower panel) plotted as a function of 
dimensionless time $t/\tau$. The coupling to the auxiliary state is $\eta = 4.0$. Note that 
since the plots show the populations of the computational, excited, and auxiliary states, 
phases cannot be seen.}
\label{fig:popH}
\end{figure}

\begin{figure}[h!]
\centering
\includegraphics[width=0.45\textwidth]{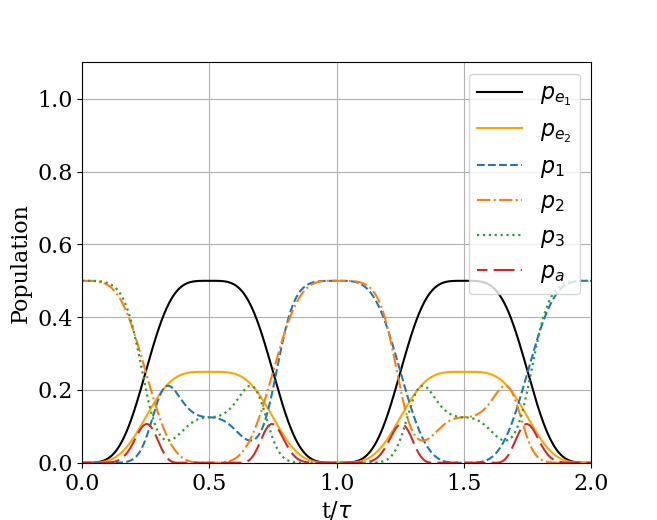}
\includegraphics[width=0.45\textwidth]{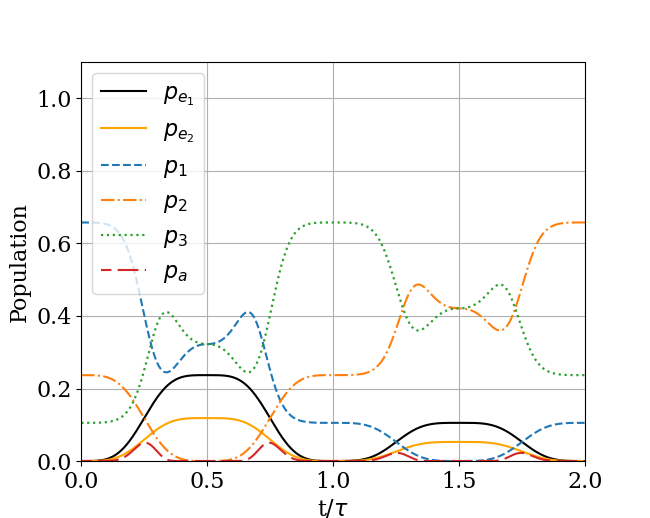}
\caption{The effect of the ${\rm X}_3$-gate on the initial states $\frac{1}{\sqrt{2}} \big(  
\ket{2} + \ket{3} \big)$ (upper panel) and $\frac{1}{\sqrt{38}} \big( 5\ket{1} + 
3\ket{2} + 2\ket{3} \big)$ (lower panel) plotted as a function of dimensionless time $t/\tau$. 
The coupling to the auxiliary state 
is $\eta = 4.0$. Note that since the plots show the populations of the computational, excited, 
and auxiliary states, phases cannot be seen.}
\label{fig:popX}
\end{figure}

The holonomy ${\rm U}_3^{(1)}$ can be parameterized by 
$\chi,\xi,\theta,\varphi, \gamma_1,\gamma_2$; 
however, these parameters are not enough to construct all gates. For instance, ${\rm X}_3$ 
requires two loops. The full gate is given by repeating ${\rm U}_3^{(1)}$ with a different set 
of parameters
\begin{eqnarray}
\label{eq:trit-gate-2-loop}
{\rm U}_3^{(1)} = {\rm U}_3^{(1)} (\chi',\xi',\theta',\varphi',\gamma_1',\gamma_2') 
{\rm U}_3^{(1)} (\chi,\xi,\theta,\varphi,\gamma_1,\gamma_2).
\end{eqnarray}
In this way, the following gates can be implemented: 
\begin{eqnarray}
{\rm X}_3 &=& U(0,0,\frac{\pi}{4},\frac{\pi}{2},0,\pi)\times U(0,0,\frac{\pi}{2},\frac{\pi}{4},0,\pi) 
\nonumber\\&=& \begin{pmatrix}
0&0&1
\\
1&0&0
\\
0&1&0
\end{pmatrix},
\nonumber \\ 
{\rm Z}_3 &=& U(0,0,0,0,\frac{2\pi}{3},\frac{4\pi}{3})
\nonumber\\&=& \begin{pmatrix}
1&0&0
\\
0&e^{i\frac{2\pi}{3}}&0
\\
0&0&e^{i\frac{4\pi}{3}}
\end{pmatrix},
\nonumber \\
{\rm T}_3 &=& U(0,0,0,0,\frac{2\pi}{9},\frac{-2\pi}{9})
\nonumber\\&=& \begin{pmatrix}
1&0&0
\\
0&e^{i\frac{2\pi}{9}}&0
\\
0&0&e^{-i\frac{2\pi}{9}}
\end{pmatrix},
\nonumber\\
{\rm H}_3 &=& U(6.41\cdot 10^{-4}, 6.56\cdot 10^{-4}, 0.48, 0.79, 1.58, 1.56) 
\nonumber\\ 
 & \times & U(9.81\cdot 10^{-3}, 0.00, 1.187, 2.15, 0.00, 1.57)
\nonumber\\&\approx & \frac{1}{\sqrt{3}}
\begin{pmatrix}
1&1&1
\\
1&e^{i\frac{2\pi}{3}}&e^{i\frac{4\pi}{3}}
\\
1&e^{i\frac{4\pi}{3}}&e^{i\frac{2\pi}{3}}
\end{pmatrix}.
\end{eqnarray}
The set includes qutrit equivalents of the Hadamard and ${\rm T}$ gate, which constitutes a 
universal set. Thus, no more than two loops for each gate are needed to achieve single qutrit 
universality. Figures \ref{fig:popH} and \ref{fig:popX} show the population of the computational, 
excited, and auxiliary states during implementation of ${\rm H}_3$ and ${\rm X}_3$, respectively, 
plotted against the dimensionless time $t/\tau$. To this end, a linear combination of $\ket{D_1(t)}$, 
$\ket{D_2(t)}$, and $\ket{\mathcal{D}}$ is matched at $t=0$ to a given initial state and the 
populations are thereafter identified by monitoring the evolving state at $t>0$. We have chosen 
$\eta = 4.0$ in the simulations to allow for direct comparison with the qubit case analyzed 
in Ref.~\cite{ai22}.

All diagonal gates can be parameterized by a single loop by fixing $\theta = \varphi = \chi = \xi = 0$. 
The dark-bright basis states reduce to $\ket{\mathcal{D}} = \ket{1}, \ket{b_1} = 
\ket{2}$, and  $\ket{b_2} = - \ket{3}$. By Eq.~(\ref{eq:trit-gate-1-loop}), one can thus see that 
all diagonal unitaries can be specified by $\gamma_1$ and $\gamma_2$: 
\begin{eqnarray}
{\rm U}_3^{(1)} (0,0,0,0,\gamma_1,\gamma_2) = \begin{pmatrix}
1&0&0
\\
0&e^{i\gamma_1}&0
\\
0&0&e^{i\gamma_2}
\end{pmatrix}.
\end{eqnarray}

\subsection{Robustness test}
We quantify gate robustness by means of fidelity 
\begin{eqnarray}
F(\psi,\tilde{\psi}) = \big| \langle \psi \ket{\tilde{\psi}} \big| 
\end{eqnarray}
with $\psi$ and $\tilde{\psi}$ the ideal and non-ideal output states of the gate, given the 
same input. The fidelity is averaged by sampling initial states and letting them evolve 
with time by numerically solving the Schrödinger equation using the SciPy implementation 
of backwards differentiation \cite{shampine97}. 

\begin{figure}[h!]
\centering
\includegraphics[width=0.5\textwidth]{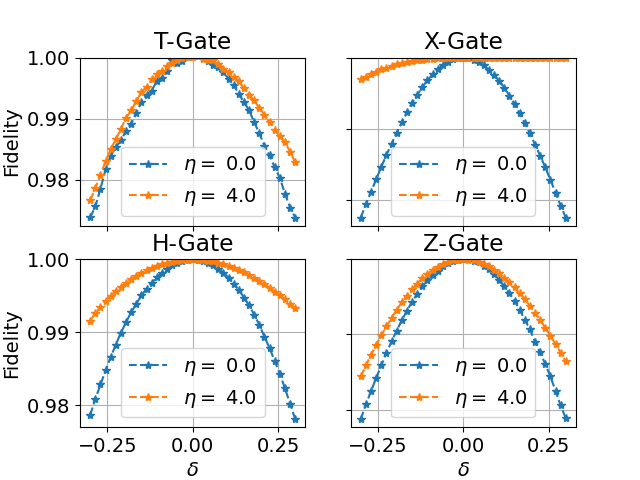}
\caption{Robustness test, average fidelity of the ${\rm T}_3,{\rm X}_3,{\rm H}_3$, and 
${\rm Z}_3$ gates. The averages are calculated by sampling over 500 randomized initial 
states with a perturbation $\Omega \mapsto (1+\delta) \Omega$ of the Rabi frequency.}
\label{fig:fidelity}
\end{figure}

We introduce systematic errors by shifting 
the Rabi frequencies $\Omega_p \mapsto \Omega_p (1 + \delta)$ for $p=1,2,a$ in 
Eq.~(\ref{eq:Ham_d}) and compare to the exact solution obtained by applying the ideal 
gate to the initial state. The calculated fidelities are shown in Fig.~\ref{fig:fidelity}. In the 
plots, it can be seen that coupling to the auxiliary state (again with $\eta = 4.0$) improves 
the robustness to Rabi frequency errors  compared to the standard non-adiabatic holonomic 
scheme ($\eta = 0$). This result is similar to that found in Ref.~\cite{ai22} for the qubit 
case, and it is reasonable to expect that it applies for higher $d$ as well. As qudits offer 
enlarged encoding space and improved  gate efficiency \cite{wang20}, this result 
demonstrates potential for dark path holonomic qudit computation. 

\section{Qudit generalization} 
\label{sec:gen}
To generalize the dark path scheme to arbitrary dimension $d$, we extend the above 
qutrit scheme by using $d$ ground states $\ket{k}$, $d-1$ excited states $\ket{e_l}$, and an 
auxiliary state $\ket{a}$. The dark path Hamiltonian   
\begin{eqnarray}\label{eq:HamN}
H^{(d)} = \sum_{k = 1}^d \sum_{l = 1}^{d-1} \omega_{k,l} \ket{k} \bra{e_l} + 
\frac{\Omega_a(t)}{2} \ket{a} \bra{e_{d-1}} + {\rm H.c.}
\label{eq:quditH}
\end{eqnarray}
is a direct extension of Eq.~(\ref{eq:Ham}). As before, the relation between the 
number of excited states and qudit ground states is chosen so as to define a single 
fixed dark eigenstate $\ket{\mathcal{D}}$. The coupling structure has the following pattern: 
the $l$th excited state $\ket{e_l}$ is connected to ground states $\ket{1},\ket{2},\ldots, 
\ket{l+1}$, except the one with the largest index $l = d-1$, which is connected to all ground 
states and the auxiliary state $\ket{a}$. Thus, $\omega_{k>l+1,l} = 0$. Just as in the qutrit 
case discussed above, dipole transitions between hyperfine levels in trapped ions is the 
ideal platform for implementing $H^{(d)}$.

Let us write the dark state fully residing in the computational subspace as 
\begin{eqnarray}
\ket{\mathcal{D}} = c_1\ket{1} + c_2\ket{2} + c_3\ket{3} \dots + c_{d}\ket{d} . 
\end{eqnarray}
This defines $d-1$ bright states $\ket{b_k}$ of the form 
\begin{eqnarray}
\label{eq:birght_states}
\ket{b_1} & = & \frac{1}{\left| c_1 \right|^2 + \left| c_2 \right|^2}
\big(-c_2^{*}\ket{1} + c_1^{*}\ket{2} \big) , 
\nonumber \\ 
\ket{b_2} & = & N_2 \big(c_1\ket{1} + c_2\ket{2} + \Lambda_3\ket{3}  \big),
\nonumber\\
&\vdots &
\nonumber\\
\ket{b_{d-1}} &=& N_{d-1} \big( c_1\ket{1} + \dots + c_{d-1} \ket{d-1} +
 \Lambda_{d} \ket{d} \big).
\end{eqnarray}
with $N_2, \ldots N_{d-1}$ being normalization factors. By construction, $\ket{b_1}$ is 
orthogonal to the dark state and all other bright states. For $k \geq 2$, $\ket{b_k}$ 
contains $k+1$ basis vectors, where the coefficient $\Lambda_{k+1}$ is chosen such 
that $\ket{b_k}$ is orthogonal to $\ket{\mathcal{D}}$. This in turn makes 
any $\ket{b_{l>k}}$ orthogonal to $\ket{b_k}$ as they have the same states and coefficients 
as $\ket{\mathcal{D}}$ for all the states involved in the inner product, which implies that 
$\langle b_{l>k} \ket{b_k} \propto \langle \mathcal{D} \ket{b_k}$. Therefore, by choosing 
the $\Lambda$'s such that these inner products are zero, the construction ensures an 
orthonormal dark-bright basis spanning the qudit subspace. Explicitly, for 
$k \geq 2$, one finds  
\begin{eqnarray}
\Lambda_{k+1} = -\frac{1}{c_{k+1}^{*}}\sum_{l = 1}^{k}|c_l|^2
\end{eqnarray}
and 
\begin{eqnarray}
N_k & = & \left( \sum_{l = 1}^k | c_l |^2  + \left| \Lambda_{k+1}\right|^2 \right)^{-1/2} 
\nonumber \\ 
 & = & \left( \sum_{l = 1}^k|c_l|^2  + \frac{1}{ \left| c_{k+1}  \right|^2}  
\left( \sum_{l = 1}^{k} |c_l|^2 \right)^2 \right)^{-1/2}.
\end{eqnarray}

\begin{figure}[h!]
\centering
\includegraphics[width=0.45\textwidth]{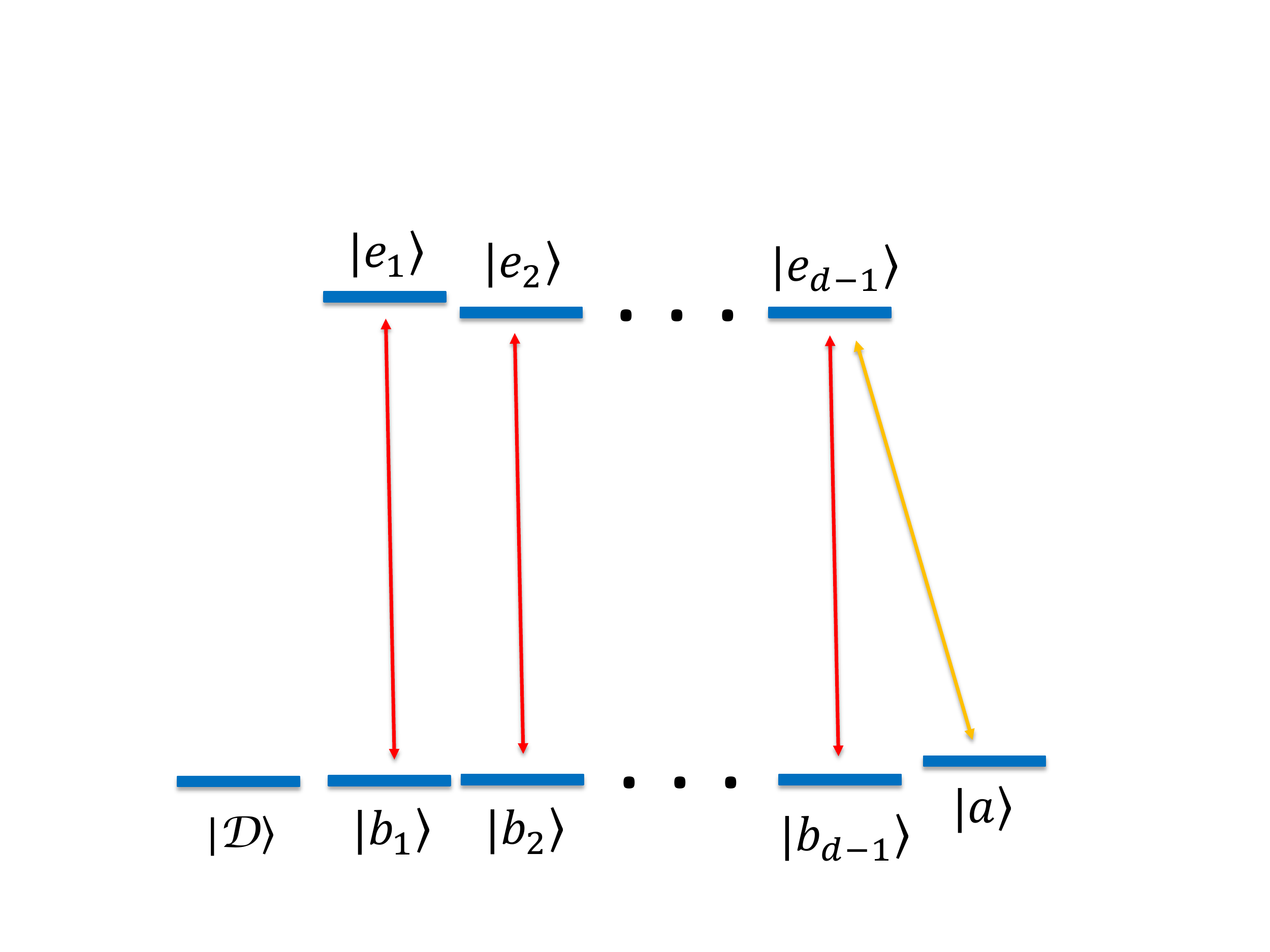}
\caption{Qudit setting in the Morris-Shore basis \cite{morris83} applied to the qudit levels 
$\ket{1}, \ldots ,\ket{d}$. As in the qutrit case, the auxiliary state $\ket{a}$ is  untouched 
and a single dark state $\ket{\mathcal{D}}$ emerges in the computational qudit subspace
${\rm Span} \{ \ket{1}, \ldots , \ket{d} \}$.}
\label{fig:setting2}
\end{figure}

In the dark-bright basis, the Hamiltonian can be written as 
\begin{eqnarray}
\label{eq:HamdN}
H^{(d)} = \sum_{k=1}^{d-1} \frac{\Omega_k(t)}{2}e^{-i\phi_k}\ket{b_k}\bra{e_k} + 
\frac{\Omega_a(t)}{2}\ket{a}\bra{e_{d-1}} + {\rm H.c.}
\end{eqnarray}
with $\Omega_k$ being real-valued time dependent Rabi frequencies and $\phi_k$ time 
independent phases. This form of $H^{(d)}$ is depicted in Fig.~\ref{fig:setting2} and 
can be used to define $d-1$ independent dark paths $\ket{D_k (t)}$; as above, these 
states must satisfy $\bra{D_k (t)}H_d\ket{D_k (t)} = 0, k = 1,\dots,d-1$, and 
$\langle D_k (t) \ket{D_l (t)} = \delta_{kl}$. The dark paths are traced out 
by the states 
\begin{eqnarray}
\ket{D_k (t)} & = & \cos u(t) e^{-i\phi_k}\ket{b_k} + i\sin u(t) \ket{e_k}, 
\nonumber \\  
k & = & 1,\ldots,d-2,
\nonumber \\
\ket{D_{d-1}(t)} & = & \cos u(t) \cos v(t) e^{-i\phi_{d-1}} \ket{b_{d-1}} - i\sin u(t) \ket{e_{d-1}} 
\nonumber \\ 
 & & - \cos u(t) \sin v \ket{a} , 
\end{eqnarray}
each of which starts and ends in one of the bright states provided $u(\tau) = u(0) = 
v(\tau) = v(0) = 0$. By using these states, one can reverse engineer the Hamiltonian 
to determine $\Omega_k$ and $\Omega_a$, yielding 
\begin{eqnarray}
\Omega_1(t) & = & \Omega_2(t) = \ldots = \Omega_{d-2}(t) = -2\dot{u},
\nonumber\\
\Omega_{d-1} (t) &=& 2\left(\dot{v}\cot u\sin v + \dot{u}\cos v \right),
\nonumber\\
\Omega_a(t) &=& 2\left(\dot{v}\cot u\cos v - \dot{u}\sin v \right).
\label{eq:genRabi}
\end{eqnarray}

Holonomic one-qudit gates can be implemented by using the same $u(t),v(t)$ as 
in the qutrit setting and by using the single-loop multi-pulse technique \cite{herterich16}. 
This results in the gate 
\begin{eqnarray}
{\rm U}_d^{(1)}  = \ket{\mathcal{D}}\bra{\mathcal{D}} + 
\sum_{k = 1}^{d-1}e^{i\gamma_k}\ket{b_k}\bra{b_k} 
\label{eq:1quditgate}
\end{eqnarray}
acting on the qudit subspace ${\rm Span} \{ \ket{\mathcal{D}},\ket{b_1},\dots, 
\ket{b_{d-1}}\}.$ The unitary is parameterized by $3(d-1)$ parameters for $d\geq 2$, i.e., 
\begin{eqnarray}
{\rm U}_d^{(1)} = {\rm U}_d^{(1)} (\theta_1,\dots,\theta_{d-1},
\varphi_1,\dots,\varphi_{d-1},\gamma_1,\dots,\gamma_{d-1}) , 
\end{eqnarray} 
where we have assumed the $c_k$'s are parameterized by the Euclidean components 
of the radius of the unit $(d-1)$-sphere with added phase factors: 
\begin{eqnarray}
c_1 &=& \cos \theta_1,
\nonumber\\ 
c_2 &=& e^{i\varphi_1}\sin \theta_1 \cos \theta_2,
\nonumber\\
&\vdots &
\nonumber\\
c_{d-1} &=& e^{i\varphi_{d-2}}\sin \theta_1 \dots \sin \theta_{d-2} \cos \theta_{d-1} ,
\nonumber\\
c_{d} &=& e^{i\varphi_{d-1}}\sin \theta_1 \dots \sin \theta_{d-2} \sin \theta_{d-1} .
\end{eqnarray}

By applying the holonomy with different parameters in sequence up to $n$ times is enough to 
create any desirable gate. To determine $n$, we use that the qudit state space is isomorphic 
to the special unitary group ${\rm SU}(d)$. By using $\dim[{\rm SU}(d)] = d^2 -1$, we deduce 
that $n$ must satisfy $3(d-1)n \geq d^2 -1$, which implies $n \geq \frac{d+1}{3}$. Thus, the 
number of loops needed to create any unitary scales linearly since some 
gates can be created with fewer loops. In particular, $n =\frac{d+1}{3}$ when $d = 3j + 2,\, j 
\in \mathbb{N}$, which are optimal qudit dimensions in the sense that they require the smallest 
number of loops per dimension and these qudits could therefore be regarded as optimal 
carriers of information since the same number of loops must be carried out while higher 
dimension has higher information capacity.

Furthermore, any diagonal gate only requires one loop. Explicitly, by setting $\theta_1 = \dots = 
\theta_{d-1} = \varphi_1 = \dots = \varphi_{d-1} = 0$ the unitary reduces to the form
\begin{eqnarray}
\label{eq:diagate}
{\rm U}_d^{(1)}(0,\dots,0,\gamma_1,\dots,\gamma_{d-1}) = \ket{1}\bra{1} + \sum_{k = 2}^d 
e^{i\gamma_k}\ket{k}\bra{k}.
\end{eqnarray}
This corresponds to the choice $c_k = \delta_{k1}$. 

\section{Holonomic dark path two-qudit gates}
\label{sec:2-qudit}
To complete the set of universal gates, we demonstrate a conditional dark path based 
holonomic gate that can entangle pairs of qudits. The scheme is adapted to trapped ions 
with their internal states encoding qudits interacting via the vibrations in the harmonic 
trap. Similar schemes have been developed for standard non-adiabatic holonomic 
quantum computation \cite{zhao19,xu21b}. 

A conditional qudit gate has the generic form 
\begin{eqnarray}
{\rm U}_d^{(2)} = \left( \begin{array}{cc} 
I_{d^2-d} & \\ 
 & V_d \end{array} \right)
\end{eqnarray}
with $V_d$ a unitary acting on the target qudit provided the control qudit is in the state 
$\ket{d}$. The $(d^2-d) \times (d^2-d)$ identity matrix $I_{d^2-d}$ acts on the remaining 
states of the qudit pair. ${\rm U}_d^{(2)}$ can be realized by designing the effective 
Hamiltonian 
\begin{eqnarray}
\mathcal{H}_{\rm eff}^{(d)} & = & \ket{d} \bra{e_{d-1}} \otimes 
\left( \sum_{k=1}^d \sum_{l=1}^{d-1} \omega_{k,l} \ket{k} \bra{e_l} \right. 
\nonumber \\ \nonumber
 & & \left. + \frac{\Omega_a (t)}{2} \ket{a} \bra{e_{d-1}} \right) + {\rm H.c.} \\ & &= \ket{d} \bra{e_{d-1}} 
 \otimes H^{(d)} + {\rm H.c.} ,
 \label{eq:eff_2qudit}
\end{eqnarray} 
which, upon use of the multi-pulse single-loop technique \cite{herterich16} applied to the 
dark path, implies that 
\begin{eqnarray}
{\rm U}_d^{(2)} & = & \mathcal{T} e^{-i\int_0^t \mathcal{H}_{\rm eff}^{(d)} dt'} 
\nonumber \\ 
 & = &  \left( \hat{1} - \ket{d} \bra{d} \right) \otimes \hat{1} + 
\ket{d} \bra{d} \otimes {\rm U}_d^{(1)}
\end{eqnarray}
with $\mathcal{T}$ time ordering and ${\rm U}_d^{(1)}$ given by Eq.~(\ref{eq:1quditgate}). 

The effective Hamiltonian in Eq.~(\ref{eq:eff_2qudit}) can be implemented in a 
S{\o}rensen-M{\o}lmer-type 
\cite{sorensen99} setup, where the single-ion transitions are  driven by two-color lasers, 
all with the same detuning $\Delta$ to the red and blue of the corresponding resonance 
frequency. By denoting the single-ion Rabi frequencies as $\omega_0$ for the control 
ion and as $\omega_1,\ldots , \omega_N,\omega_a$, $N=\frac{1}{2}(d^2+d)-1$, 
for the target ion, we obtain the Hamiltonian
\begin{eqnarray}
\mathcal{H}^{(d)} & = & i\eta_L \left( b e^{-i\nu t} + b^{\dagger} e^{i\nu t} \right) \otimes \Big[ 
\omega_0 (t) \ket{d} \bra{e_{d-1}} \otimes \hat{1} 
\nonumber \\ 
 & & + \hat{1} \otimes \Big( \omega_1 \ket{1} \bra{e_1} + \omega_2 \ket{2} \bra{e_1} + 
\omega_3 \ket{1} \bra{e_2} + 
\ldots 
\nonumber \\ 
 & & \ldots + \omega_N \ket{d} \bra{e_{d-1}} + 
\omega_a \ket{a} \bra{e_{d-1}} \Big)  \Big] \cos \Delta t
\nonumber \\ 
 & & + {\rm H.c.}  
\end{eqnarray} 
responsible for the qudit-qudit interaction. Here, $b$ ($b^{\dagger}$) is the annihilation 
(creation) operator of the vibrational mode with frequency $\nu$ and $\eta_L$ is the Lamb-Dicke 
parameter satisfying the Lamb-Dicke criterion $\eta_L \ll 1$. In the large detuning limit, single-ion 
transitions are strongly suppressed and one may use the technique developed in 
Ref.~\cite{james07} to derive  the effective Hamiltonian 
\begin{eqnarray}
\widetilde{\mathcal{H}}_{\rm eff}^{(d)} =  \mathcal{H}_{\rm eff}^{(d)} + \bar{\mathcal{H}}_{\rm eff}^{(d)}
\end{eqnarray}
with 
\begin{eqnarray}
\bar{\mathcal{H}}_{\rm eff}^{(d)} & = & - \ket{d} \bra{e_{d-1}} \otimes 
\left( \sum_{k=1}^d \sum_{l=1}^{d-1} \omega_{k,l}^{\ast} \ket{e_l} \bra{k} \right. 
\nonumber \\ 
 & & \left. + \frac{\Omega_a^{\ast} (t)}{2} \ket{e_{d-1}} \bra{a} \right) + {\rm H.c.} 
\end{eqnarray} 
up to Stark shift contributions that can be compensated for by applying additional laser 
pulses. Here, 
\begin{eqnarray}
\omega_{1,1} & = & k \left| \omega_0 \omega_1 \right| e^{i\phi_1}, \omega_{1,2} = 
k \left| \omega_0 \omega_2 \right| e^{i\phi_2}, \ldots , 
\nonumber \\ 
\omega_{d,d-1} & = &  
k \left| \omega_0 \omega_N \right| e^{i\phi_N}, 
\Omega_a = 2k \left| \omega_0 \omega_a \right| e^{i\phi_a} 
\end{eqnarray} 
with 
\begin{eqnarray}
k & = & \eta_L^2 \frac{\nu}{\Delta^2 - \nu^2} , 
\nonumber \\ 
\omega_j & = & |\omega_j| e^{i(\phi_j-\phi_0)}, \ \ j=0,1,\ldots,N,a . 
\end{eqnarray}
We note that $\bar{\mathcal{H}}_{\rm eff}^{(d)}$ 
commutes with $\mathcal{H}_{\rm eff}^{(d)}$, which implies  
\begin{eqnarray}
\mathcal{T} e^{-i\int_0^t \widetilde{\mathcal{H}}_{\rm eff}^{(d)} dt'} = \mathcal{T} 
e^{-i\int_0^t \mathcal{H}_{\rm eff}^{(d)} dt'} 
\mathcal{T} e^{-i\int_0^t \bar{\mathcal{H}}_{\rm eff}^{(d)} dt'} . 
\label{eq:reduction}
\end{eqnarray}
The second factor of the right-hand side of Eq.~(\ref{eq:reduction}) 
acts trivially on the computational subspace ${\rm Span} \{ \ket{kl}, k,l=1,\ldots,d \}$, 
thus effectively reducing $\widetilde{\mathcal{H}}_{\rm eff}^{(d)}$ to 
$\mathcal{H}_{\rm eff}^{(d)}$. 

As a final remark, up to the laser pulses needed to compensate for Stark shifts, there 
is only one additional laser needed to implement ${\rm U}_d^{(2)}$ as compared to 
${\rm U}_d^{(1)}$. Thus, the resources to implement the one- and two-qudit holonomic 
gates are roughly the same. 

\section{Conclusions}
We have shown how to explicitly create a quantum mechanical system, which could be used 
to emulate a qutrit and corresponding universal set of one-qutrit holonomic gates. This is done by 
expanding the dark path qubit scheme \cite{ai22} into three dimensions. We have shown 
how it generalizes in the one- and two-qudit case and how the use of auxiliary states can 
improve the robustness of the gates. 

The qutrit gates have high fidelity and their robustness is improved by the inclusion of the 
auxiliary state in a similar way as for the qubit, which suggests that the dark path method can be 
beneficial for higher dimensional qudits to improve robustness. In the general qudit case, 
we have shown how any one-qudit diagonal unitary could be created by a single multi-pulse 
loop in parameter space and that non-diagonal unitaries scale linearly in the number 
of loops required for control of each loop. The possibility that the scheme expands 
efficiently into certain dimensions has been discussed.

\section*{Acknowledgment}
E.S. acknowledges financial support from the Swedish Research Council (VR) through 
Grant No. 2017-03832.


\begin{thebibliography}{99}
\bibitem{zanardi99} P. Zanardi and M. Rasetti,  
Holonomic quantum computation, 
Phys. Lett. A {\bf 264}, 94 (1999). 
\bibitem{sjoqvist12} E. Sj\"{o}qvist, D. M. Tong, L. M. Andersson, B. Hessmo, M. Johansson, 
and K. Singh, 
Non-adiabatic holonomic quantum computation, 
New J. Phys. {\bf 14}, 103035 (2012).
\bibitem{wilczek84} F. Wilczek and A. Zee, 
Appearance of Gauge Structure in Simple Dynamical Systems, 
Phys. Rev. Lett. {\bf 52}, 2111  (1984).
\bibitem{anandan88} J. Anandan, 
Non-adiabatic non-Abelian geometric phase, 
Phys. Lett. A {\bf 133}, 171 (1988). 
\bibitem{zhang18} J. Zhang, S. J. Devitt, J. Q. You, and F. Nori, 
Holonomic surface codes for fault-tolerant quantum computation, 
Phys. Rev. A {\bf 97}, 022335 (2018). 
\bibitem{xu18a} G. F. Xu, D. M. Tong, and E. Sj\"oqvist, 
Path-shortening realizations of nonadiabatic holonomic gates, 
Phys. Rev. A {\bf 98}, 052315 (2018).
\bibitem{liu19} B.-J. Liu, X.-K. Song, Z.-Y. Xue, X. Wang, and M.-H. Yung, 
Plug-and-Play Approach to Nonadiabatic Geometric Quantum Gates,
Phys. Rev. Lett. {\bf 123}, 100501 (2019). 
\bibitem{li20} S. Li, T. Chen, and Z.-Y. Xue, 
Fast holonomic quantum computation on superconducting circuits with optimal control, 
Adv. Quantum Technol. {\bf 3}, 2000001 (2020). 
\bibitem{dong21a} W. Dong, F. Zhuang, S. E. Economou, and E. Barnes, 
Doubly geometric quantum control, 
PRX Quantum {\bf 2}, 030333 (2021). 
\bibitem{chen21} Y.-H. Chen, W. Qin, R. Stassi, X. Wang, and F. Nori, 
Fast binomial-code holonomic quantum computation with ultrastrong light-matter coupling, 
Phys. Rev. Research {\bf 3}, 033275 (2021).
\bibitem{dong21b} Y. Dong, C. Feng, Y. Zheng, X.-D. Chen, G.-C. Guo, and F.-W. Sun, 
Fast high-fidelity geometric quantum control with quantum brachistochrones,
Phys. Rev. Research {\bf 3}, 043177 (2021).
\bibitem{setiawan21} F. Setiawan, P. Groszkowski, H. Ribeiro, and A. A. Clerk, 
Analytic design of accelerated adiabatic gates in realistic qubits: General theory 
and applications to superconducting circuits, 
PRX Quantum {\bf 2}, 030306 (2021).
\bibitem{alves22} G. O. Alves and E. Sj\"oqvist, 
Time-optimal holonomic quantum computation, 
Phys. Rev. A {\bf 106}, 032406 (2022). 
\bibitem{xu18b} Y. Xu, W. Cai, Y. Ma, X. Mu, L. Hu, T. Chen, H. Wang, Y. P. Song, 
Z.-Y. Xue, Z.-q. Yin, and L. Sun, 
Single-Loop Realization of Arbitrary Nonadiabatic Holonomic Single-Qubit Quantum Gates 
in a Superconducting Circuit, 
Phys. Rev. Lett. {\bf 121}, 110501 (2018). 
\bibitem{huang19} Y.-Y. Huang, Y.-K. Wu, F. Wang, P.-Y. Hou, W.-B. Wang, W.-G. Zhang, 
W.-Q. Lian, Y.-Q. Liu, H.-Y. Wang, H.-Y. Zhang, L. He, X.-Y. Chang, Y. Xu, and L.-M. Duan, 
Experimental Realization of Robust Geometric Quantum Gates with Solid-State Spins, 
Phys. Rev. Lett. {\bf 122}, 010503 (2019). 
\bibitem{egger19} D. J. Egger, M. Ganzhorn, G. Salis, A. Fuhrer, P. Müller, P. Kl. Barkoutsos, 
N. Moll, I. Tavernelli, and S. Filipp, 
Entanglement Generation in Superconducting Qubits Using Holonomic Operations, 
Phys. Rev. Appl. {\bf 11}, 014017 (2019).
\bibitem{xu20} Y. Xu, Z. Hua, Tao Chen, X. Pan, X. Li, J. Han, W. Cai, Y. Ma, H. Wang, 
Y. P. Song, Z.-Y. Xue, and L. Sun, 
Experimental Implementation of Universal Nonadiabatic Geometric Quantum Gates in a 
Superconducting Circuit, 
Phys. Rev. Lett. {\bf 124}, 230503 (2020). 
\bibitem{dong21c} Y. Dong, S.-C. Zhang, Y. Zheng, H.-B. Lin, L.-K. Shan, X.-D. Chen, 
W. Zhu, G.-Z. Wang, G.-C. Guo, and F.-W. Sun, 
Experimental Implementation of Universal Holonomic Quantum Computation on Solid-State 
Spins with Optimal Control, 
Phys. Rev. Appl. {\bf 16}, 024060 (2021).
\bibitem{xu21a} K. Xu, W. Ning, X.-J. Huang, P.-R. Han, H. Li, Z.-B. Yang, D. Zheng, H. Fan, 
and S.-B. Zheng, 
Demonstration of a non-Abelian geometric controlled-NOT gate in a superconducting circuit, 
Optica {\bf 8}, 972 (2021).
\bibitem{wu22} J.-L. Wu,  S. Tang, Y. Wang, X.-S. Wang, J.-X. Han, C. L\"u, J. Song, 
S.L. Su, Y. Xia, and Y.-Y. Jiang, 
Unidirectional acoustic metamaterials based on nonadiabatic holonomic quantum transformations, 
Sci. China: Phys. Mech. Astron. {\bf 65}, 220311 (2022). 
\bibitem{ai22} M.-Z. Ai, S. Li, R. He, Z.-Y. Xue, J.-M. Cui, Y.-F. Huang, C.-F. Li and G.-C. Guo, 
Experimental realization of nonadiabatic holonomic single-qubit quantum gates with two 
dark paths in a trapped ion, 
Fundam. Res. {\bf 2}, 661 (2022). 
\bibitem{glusker05} M. Glusker, D. M. Hogan and P. Vass, 
The ternary calculating machine of Thomas Fowler, 
IEEE Ann. Hist. Comput. {\bf 27}, 4 (2005).
\bibitem{bursentsov11} N. P. Brusentsov and J. R. Alvarez, 
Ternary computers: The Setun and the Setun 70, 
in Perspectives on Soviet and Russian 
Computing, edited by J. Impagliazzo and E. Proydakov, {\bf 357}, 74 (2011). 
\bibitem{lanyon09} B. P. Lanyon, M. Barbieri, M. P. Almeida, T. Jennewein, T. C. Ralph, 
K. J. Resch, G. J. Pryde, J. L. O'Brien, A. Gilchrist, and A. G. White,  
Simplifying quantum logic using higher-dimensional Hilbert spaces, 
Nat. Phys. {\bf 5}, 134 (2009). 
\bibitem{baekkegaard19} T. Baekkegaard, L. B. Kristensen, N. J. S. Loft, C. K. Andersen, 
D. Petrosyan, and N. T. Zinner, 
Realization of efficient quantum gates with a superconducting qubit-qutrit circuit, 
Sci. Rep. {\bf 9}, 13389 (2019). 
\bibitem{chi22} Y. Chi, J. Huang, Z. Zhang, J. Mao, Z. Zhou, X. Chen, C. Zhai, J. Bao, 
T. Dai, H. Yuan, M. Zhang, D. Dai, B. Tang, Y. Yang, Z. Li, Y. Ding, L. K. Oxenl{\o}we, 
M. G. Thompson, J. L. O'Brien, Y. Li, Q. Gong, and J. Wang, 
A programmable qudit-based quantum processor, 
Nat. Commun. {\bf 13}, 1166 (2022). 
\bibitem{wang20} Y. Wang, Z. Hu, B. C. Sanders, and S. Kais,  
Qudits and high-dimensional quantum computing, 
Front. Phys. {\bf 8}, 589504 (2020). 
\bibitem{shkolnikov20} V. O. Shkolnikov and G. Burkard, 
Effective Hamiltonian theory of the geometric evolution of quantum systems, 
Phys. Rev. A {\bf 101}, 042101 (2020). 
\bibitem{low20} P. J. Low, B. M. White, A. A. Cox, M. L. Day, and C. Senko, 
Practical trapped-ion protocols for universal qudit-based quantum computing, 
Phys. Rev. Research {\bf 2}, 033128 (2020). 
\bibitem{morris83} J. R. Morris and B. W. Shore, 
Reduction of degenerate two-level excitation to independent two-state system, 
Phys. Rev. A {\bf 27}, 906 (1983).
\bibitem{herterich16} E. Herterich and E. Sj\"oqvist,  
Single-loop multiple-pulse nonadiabatic holonomic quantum gates, 
Phys. Rev. A {\bf 94}, 052310 (2016). 
\bibitem{shampine97} L. F. Shampine and M. W. Reichelt, 
The MATLAB ODE Suite, 
SIAM J. Sci Comput. {\bf 18}, 1 (1997).
\bibitem{zhao19} P. Z. Zhao, G. F. Xu, and D. M. Tong, 
Nonadiabatic holonomic multiqubit controlled gates, 
Phys. Rev. A {\bf 99}, 052309 (2019). 
\bibitem{xu21b} G. F. Xu, P. Z. Zhao, E. Sj\"oqvist, and D. M. Tong, 
Realizing nonadiabatic holonomic quantum computation beyond the three-level setting, 
Phys. Rev. A {\bf 103}, 052605 (2021). 
\bibitem{sorensen99} A. S{\o}rensen and K. M{\o}lmer, 
Quantum Computation with Ions in Thermal Motion, 
Phys. Rev. Lett. {\bf 82}, 1971 (1999). 
\bibitem{james07} D. F. V. James and J. Jerke, 
Effective Hamiltonian theory and its applications in quantum information, 
Can. J. Phys. {\bf 85}, 625 (2007).
\end{thebibliography}
\end{document}